\documentclass[amsmath,amssymb,aps,reprint]{revtex4-1}

\newcommand{\be}{\begin{equation}}
\newcommand{\ee}{\end{equation}}
\newcommand{\bea}{\begin{eqnarray}}
\newcommand{\eea}{\end{eqnarray}}
\newcommand{\nnn}{\nonumber}

\begin{document}

\title{Thermodynamic Interpretation of Soft Glassy Rheology Models}

\author{Peter Sollich}%
\email{peter.sollich@kcl.ac.uk}
\affiliation{King's College London, Department of Mathematics, Strand,
  London WC2R 2LS, UK}
\author{Michael E. Cates}
\affiliation{SUPA, School of Physics and Astronomy, The University of Edinburgh, \\JCMB, The King's Buildings, Mayfield Road, Edinburgh EH9 3JZ, UK}

\date{15 Jan 2012}

\begin{abstract}
Mesoscopic models play an important role in our understanding of the deformation and flow of amorphous materials. One such description, based on the Shear Transformation Zone (STZ) theory, has recently been re-formulated within a non-equilibrium thermodynamics framework, and found to be consistent with it. We show here that a similar interpretation can be made for the Soft Glassy Rheology (SGR) model. Conceptually this means that the ``noise temperature" $x$, proposed phenomenologically in the SGR model to control the dynamics of a set of slow mesoscopic degrees of freedom,
can consistently be interpreted as their actual thermodynamic
temperature. (Because such modes are slow to equilibrate, this generally does not coincide with the temperature of the fast degrees of freedom and/or heat bath.) If one chooses to make this interpretation, the thermodynamic framework significantly constrains extensions of the SGR approach to models in which $x$ is a dynamical variable. We assess in this light some such extensions recently proposed in the context of shear banding.
\end{abstract}

\pacs{to be added}

\maketitle

\section{Introduction}

Developing predictive theories for the deformation and flow of
amorphous materials remains an ongoing challenge. The class of
materials concerned is large, including not only colloidal glasses,
emulsions and foams but also
molecular glass formers, metallic glasses and possibly granular materials.
The development of such predictive theories would therefore be of
great practical significance as well as fundamental theoretical
interest~\cite{review_glasses}.
Two key challenges in understanding deformation and flow in
amorphous materials are the absence of a reference configuration
relative to which strain can be defined (in contrast to ordered crystalline solids), and the difficulty of encapsulating local flow events and/or larger-scale stress redistributions in a coarse-grained variable or
continuum field (analogous to the dislocation density for crystals).

First principles approaches to glass rheology have made significant progress~\cite{BraVoiCatFuc07,BraCatFuc08}, but entail major approximations (such as mode coupling theory), usually followed by further simplification~\cite{BraVoiFucLarCat09}.
In the absence of a comprehensive microscopic theory, a number of models have been developed directly at a mesoscopic level to
describe amorphous
flow, see e.g.~\cite{HebLeq98,PicAjdBocLeq02,BocColAjd09,PouFor09,TalPetVanRou11,ManColChaBoc11}. Two such
approaches that have been investigated in some detail are the
shear transformation zone (STZ) approach, reviewed recently
in~\cite{FalLan11}, and the soft glassy rheology (SGR)
model~\cite{SolLeqHebCat97,Sollich98,FieSolCat00,FieCatSol09}. One important
feature of the STZ theory is that it can be given an unambiguous thermodynamic interpretation,
as discussed in detail in a series of recent
papers~\cite{BouLan09,BouLan09b,BouLan09c}. 

Our purpose in this paper
is to show 
that the SGR model can likewise be cast consistently within a non-equilibrium thermodynamic
framework. This has two main benefits. Firstly, it gives an
alternative way of looking at the effective temperature parameter $x$ that was previously proposed within SGR to set the noise level for stochastic mesoscopic dynamics.  This parameter can now, if desired, by interpreted as a genuine non-equilibrium thermodynamic temperature, governing a subset of degrees of freedom whose dynamics causes the system to move among its various ``inherent structures" or energy minima. Secondly, this interpretation significantly constrains how SGR can be extended to allow $x$ to evolve in time and space: it suggests a specific form for the driving term in the dynamical equation for an evolving $x$. We find that this driving
term is consistent, in steady state, with one of two extensions of SGR postulated recently
to understand shear banding in amorphous flows~\cite{FieCatSol09}, but not with the other. Similar remarks apply to two models for shear-thickening fluids developed previously~\cite{HeaAjdCat01}.

We begin by reviewing briefly the SGR model, and then the
non-equilibrium thermodynamic framework of Bouchbinder and
Langer~\cite{BouLan09,BouLan09b,BouLan09c}
(Secs.~\ref{background:SGR} and~\ref{background:thermo}). Sec.~\ref{consistency} has the core of our
argument, which shows that in this framework SGR can indeed be written
as a thermodynamically consistent model, and that this consistency
forces $x$ to be the non-equilibrium thermodynamic temperature of a set of slow degrees of freedom. In
Sec.~\ref{extensions} we discuss to what extent this argument can be
made to cover also generalizations of the SGR
model. Sec.~\ref{sec:conclusion} has a summary of our results and
discusses the implications for various extended SGR models that treat $x$ as a dynamical variable.

\section{The Soft Glassy Rheology model}
\label{background:SGR}

The SGR model~\cite{SolLeqHebCat97,Sollich98,FieSolCat00,FieCatSol09}
describes a sample of amorphous material as a collection of mesoscopic
elements, chosen large enough that it makes sense to define
local strain and stress variables. The local strain $l$ is defined
relative to a local equilibrium state~\cite{SGRmeso}, and is assumed
to change in time by following the change in the macroscopic shear
strain $\gamma$, so that $\dot l = \dot\gamma$. We focus here on pure
shear strain 
deformation but note that the model can also be extended to general deformation
tensors~\cite{CatSol04}. The local strain cannot grow indefinitely:
once the stored elastic energy, which we write as $(k v_e/2)l^2$ in
terms of the local shear modulus $k$ and the element volume $v_e$, becomes
close to a local yield energy $E$, a yield event will take place. This
is assumed to reset $l$ to zero and create a new local equilibrium
configuration with correspondingly a new yield energy, drawn from a
distribution $\rho(E)$. The final ingredient is that this yield
process is taken as activated by some effective temperature $x$. This
is based on the intuition that every yield event elsewhere in the
material causes a `kick' locally via the associated stress
redistribution, and that many such kicks add up to an effectively thermal noise.

The resulting equation of motion for the distribution of strain and
yield energy across elements, $P(E,l,t)$, is
\be
\dot P(E,l,t) = -\dot\gamma \partial_l P - \Gamma(E,l)P +
Y(t)\rho(E,l)
\label{SGR_eqn}
\ee
Here $\Gamma(E,l)$ is the yield rate from a given local configuration,
which represents the thermal activation in the form 
$\Gamma(E,l)=\Gamma_0\exp[-(E-kv_el^2/2)/x]$. (We will leave
$\Gamma(E,l)$ general whenever possible in the following, to assess
the scope of our arguments.) We use $k_{\rm B}=1$
throughout, so that all temperatures are measured in energy units. The time-dependent
total yield rate is $Y(t)=\int dE\,dl\,\Gamma(E,l) P(E,l,t)$.
Finally, $\rho(E,l)$ is the probability density of $E$ and $l$ for elements after a
jump. This is equal to $\rho(E)\delta(l)$ in the standard SGR model
as described above. By choosing $\rho(E)$ to have an exponential tail
$\sim e^{-E/E_0}$ one finds that a glass transition arises at
$x/E_0=1$.
For $x/E_0>1$ the material is ergodic (albeit with infinite viscosity for $1<x/E_0<2$) whereas for $x/E_0<1$ the material is a nonergodic amorphous solid with time-dependent (aging) material properties~\cite{FieSolCat00}.

The SGR shear stress is an average over elements,
\be
\sigma(t) = \int dE\,dl\,kl P(E,l,t)
\label{SGR_stress}
\ee
with $k$ the local shear modulus as defined above. Starting from some initial
condition $P(E,l,0)$, this together with the evolution equation for
$P(E,l,t)$ in principle determines the stress $\sigma(t)$ for any imposed shear
history $\gamma(t)$.

\section{Non-equilibrium thermodynamics framework}
\label{background:thermo}

We next review the non-equilibrium thermodynamic framework of
Bouchbinder and Langer~\cite{BouLan09,BouLan09b,BouLan09c}. The basic
premise is that the degrees of freedom of the material can be divided
up, conceptually, into two weakly interacting subsystems. The degrees
of freedom in the configurational subsystem ($C$) encode which of the
exponentially many local energy minima, known in the glass community
as inherent structures, the system finds itself in; this subsystem is
`slow' because at low temperature any motion between inherent structures
has to be thermally activated or driven by external deformation. The
kinetic-vibrational subsystem ($K$) gathers the remaining degrees of
freedom that describe fast motion around these inherent structures.
This subsystem is taken as strongly coupled to a thermal reservoir
($R$) that sets the thermodynamic equilibrium temperature $\theta_R$.
The total internal energy of the system is then written as
\[
U_{\rm tot} = U_C(S_C,\Lambda) + U_K(S_K) + U_R(S_R)
\]
where $S_C$, $S_K$ and $S_R$ are the entropies of the configurational
and kinetic-vibrational subsystems and of the thermal reservoir. In
$U_C$, $\Lambda$ denotes a set of internal state variables of the
configurational subsystem. In the STZ context these are taken as
concentrations of shear transformation zones in different
orientations. For our application to SGR, we will take $\Lambda$ as
the distribution $P(E,l)$. Note that because this distribution
already contains information on all local elastic strains, we do not include
separately in $U_C$ a state variable for strain as was done
in~\cite{BouLan09b,BouLan09c}.

The thermodynamic analysis is based on the first and second laws. The
first law, i.e.\ energy conservation, takes the following form:
\be
\dot U_C + \dot U_K + \dot U_R = V\sigma \dot\gamma
\ee
The r.h.s.\ is the external work rate that arises from
shearing a system of volume $V$ at shear rate $\dot\gamma$ against a
shear stress $\sigma$. In terms of the subsystem entropies 
$S_C$, $S_K$ and $S_R$, and the corresponding temperatures
$\chi=\left.\partial U_C/\partial S_C\right|_\Lambda$ and
$\theta=\partial U_K/\partial S_K$, the first law can be rewritten as
\be
\chi \dot S_C + \theta \dot S_K + \dot U_R - V\sigma \dot\gamma +
\dot\Lambda \left.\frac{\partial U_C}{\partial \Lambda}\right|_{S_C} =
0
\label{1st_law}
\ee
We only write one $\Lambda$-derivative term here, as a shorthand for
a sum over the relevant derivatives w.r.t.\ the state variables
gathered in $\Lambda$.

The second law states that total entropy must increase, i.e.
\be
\dot S_C + \dot S_K + \dot S_R \geq 0
\ee
One can eliminate $\dot S_C$ from this using the first law and write
$\dot S_R = \dot U_R/\theta_R$ to get
\be
-\left(1-\frac{\theta}{\theta_R}\right) \dot U_R -
\left(1-\frac{\chi}{\theta}\right) \left(\theta\dot S_K +
\frac{\theta}{\theta_R} \dot U_R\right) + W \geq 0
\label{2nd_law}
\ee
where
\be
W = 
V\sigma \dot\gamma -
\dot\Lambda \left.\frac{\partial U_C}{\partial \Lambda}\right|_{S_C}
\label{W_def}
\ee
Because time variations in $U_R$, $S_K$ and $\Lambda$ are in principle
arbitrary, it is plausible to
argue~\cite{BouLan09b} that the three terms in 
(\ref{2nd_law}) should be separately non-negative. 
To allow easier comparison with \cite{BouLan09,BouLan09b,BouLan09c}, we follow this
reasoning here. However, it turns out that this separate non-negativity is sufficient but not necessary for (\ref{2nd_law}): in Appendix \ref{appendix} we present a less restrictive but still sufficient set of conditions for  non-negative entropy production. Setting that aside and assuming separate non-negativity of all three terms, one
obtains the conditions $W\geq 0$ and
\be
\dot U_R = -B(\theta_R-\theta), 
\qquad
\theta\dot S_K +
\frac{\theta}{\theta_R} \dot U_R = -A(\theta-\chi)
\label{inequalities}
\ee
where $A$ and $B$ are positive but can depend on the state of the
system, e.g.\ via the various temperatures. Our notation for $A$
follows Ref.~\cite{FalLan11}, and we define $B$ by analogy; the coefficients
in~\cite{BouLan09,BouLan09b,BouLan09c} differ by factors of
temperature. To simplify further, one assumes strong coupling between
the thermal reservoir and the fast subsystem. In the corresponding
limit of large $B$, one then obtains $\theta=\theta_R$, but this equality is approached in such a way that
the product $B(\theta_R-\theta)$ stays finite. In fact it is only in this limit that (\ref{2nd_law}) leads to (\ref{inequalities}), as discussed further in Appendix
\ref{appendix}.

One can now write the first law as
%
%
\be
\chi \dot S_C = 
%
%
- (\theta \dot S_K + \dot U_R) + W
\ee
%
%
Because of the assumption of strong coupling between the reservoir and
the fast subsystem, one can exploit $\theta=\theta_R$ to write the
first term on the r.h.s.\ as  $-(\theta\dot S_K +
\frac{\theta}{\theta_R} \dot U_R) = A(\theta-\chi)$, so
\be
\chi \dot S_C = 
%
%
W + A(\theta-\chi) \label{eqom}
\ee
The upshot of this analysis~\cite{BouLan09b,BouLan09c} is thus
twofold. Firstly, one obtains an
equation of motion for the configurational entropy; this also then
determines the dynamics of the non-equilibrium
temperature $\chi$ as discussed in Sec.~\ref{sec:conclusion}.
The second result is the constraint $W\geq 0$. In (\ref{eqom}), $W$ is
the dissipation rate, i.e.\ the difference between the external work
(rate) and the (rate of) increase in the free energy of the
$\Lambda$-degrees of freedom.

To go beyond the formal expression for $W$ above requires further
assumptions on the relation between the state variables $\Lambda$ and
the other configurational degrees of
freedom~\cite{BouLan09,BouLan09b,BouLan09c}. The slow subsystem as a
whole is described via $U_C(S_C,\Lambda) = U_\Lambda(\Lambda) +
U_1(S_C-S_\Lambda(\Lambda))$ where $U_1(S_1)$ captures all the slow
degrees of freedom other than $\Lambda$. Then
\be
\left.
\frac{\partial U_C}{\partial\Lambda}\right|_{S_C}
= \frac{\partial U_\Lambda}{\partial \Lambda} - \frac{\partial
U_1}{\partial S_1} \frac{\partial S_\Lambda}{\partial\Lambda}
\ee
One can now simplify using $\chi=\left.\partial U_C/\partial
  S_C\right|_\Lambda = \partial U_1/\partial S_1$. Assuming further
that the number of internal state variables in $\Lambda$ is much
smaller than the overall number of slow degrees of
freedom~\cite{BouLan09}, one has $S_\Lambda\ll S_C$ and $\chi$ can be
regarded as a function of this vast majority of configurational
degrees only. The final expression for the dissipation rate is then
\be
W = V\sigma\dot\gamma - 
\dot\Lambda \left(\frac{\partial U_\Lambda}{\partial \Lambda} - \chi
\frac{\partial S_\Lambda}{\partial\Lambda}\right)
\label{W_general}
\ee
and this shows that external work that is not dissipated is indeed
stored in the free energy $U_\Lambda-\chi S_\Lambda$ of the internal
state variables $\Lambda$.

\section{Application to the SGR model}
\label{consistency}

As mentioned earlier, in applying the above formalism to SGR our
approach we take the distribution $P(E,l)$ as the set of internal state
variables $\Lambda$. The arguments above rely on the number of degrees of
freedom in $\Lambda$ being much smaller than the overall number of
degrees of freedom in the configurational subsystem. This can be
ensured by initially considering, instead of $P(E,l)$, the total probability
this distribution assigns to each of a finite number of bins covering
the $(E,l)$ 
plane. The number of bins is then sent to infinity more slowly than
the system volume $V$ in the thermodynamic limit of large $V$. We
assume below that this limit has been taken appropriately and write
directly the version of the calculations in the large system
limit, i.e.\ without binning.

Clearly for SGR one should choose as the internal energy
associated with $\Lambda \equiv P(E,l)$,
\be
U_\Lambda(\Lambda) = N_e \int dE\, dl\, P(E,l) u(E,l)
\ee
where $N_e=V/v_e$ is the number of SGR elements with volume $v_e$
each, and $u(E,l)$ is some function of the local yield energy and strain.
The corresponding choice of entropy is less obvious. A natural proposal is $S_\Lambda(\Lambda)=
-N_e \int dE\,dl\, P(E,l)[\ln P(E,l)-1]
$. However, it turns out that this choice does not give a
thermodynamically consistent interpretation of the SGR
model, essentially because every yield event resets $l=0$ and so loses an infinite
amount of entropy. We argue instead that, because the $l$-dynamics is
deterministic apart from jumps when rearrangements occur, these
degrees of freedom are ``slaved''. Hence we should only consider the
entropy of the distribution across yield energies. We also allow for a
prior distribution (or density of states) $R(E)$ in the entropy, and take:
\be
S_\Lambda(\Lambda) = -N_e \int dE\,dl\, P(E,l)\left[\ln
\frac{P(E)}{R(E)}-1\right]
\ee
where $P(E)=\int dl\, P(E,l)$. The argument that the entropy should be
independent of the $l$-degrees of freedom is somewhat analogous to
the observation made in STZ that in order to derive
the standard STZ equations of motion one has to assume that the shear
transformation zone orientation makes no contribution to the
entropy~\cite{FalLan11}.

To ensure that SGR fits into the thermodynamic framework described
above, we need to show that with the above choices for $U_\Lambda$ and
$S_\Lambda$, one has $W\geq 0$ always, which means that the rate at
which external
work is performed is never less than the rate at which free energy is stored reversibly in the internal
state variables $\Lambda\equiv P(E,l)$.

The inequality $W\geq 0$ has to hold at all times $t$, so we drop all
time arguments in the following. 
If we unpack our shorthand notation, the contribution to $W$ in
(\ref{W_general}) from the change of $P(E,l)$ becomes
\be
\dot\Lambda \frac{\partial (U_\Lambda-\chi S_\Lambda)}{\partial
  \Lambda}
\equiv \int dE\, dl\, \dot P(E,l)
\frac{\delta (U_\Lambda-\chi S_\Lambda)}{\delta P(E,l)}
\ee
where the functional derivative on the right hand side is
\be
\frac{\delta (U_\Lambda-\chi S_\Lambda)}{\delta P(E,l)}
= N_e\left[u(E,l) + \chi \ln\frac{P(E)}{R(E)}\right]
\ee
In combination with the external work, this gives for the
dissipation rate
\be
W = V\sigma\dot\gamma - 
N_e \int dE\, dl\, \dot P(E,l) 
\left[u(E,l) + \chi \ln\frac{P(E)}{R(E)}\right]
\ee
We now want to check that $W\geq 0$, for the SGR equation of motion
(\ref{SGR_eqn}). Inserting the latter into the expression above and
also expressing $\sigma$ as an average over $P(E,l)$ from
(\ref{SGR_stress}) shows
\bea
\frac{W}{N_e} &=& \int dE\,dl\,\left\{
\dot\gamma kv_e l P + \dot\gamma \partial_l P
\left[u + \chi \ln\frac{P(E)}{R(E)}\right]\right.
\nnn
\\
&&\left.{}
+(\Gamma P - Y \rho)\left[u + \chi \ln\frac{P(E)}{R(E)}\right]\right\}
\eea
where $P$, $\rho$, $u$ are short for $P(E,l)$, $\rho(E,l)$, $u(E,l)$
respectively. Now integrate by parts over $l$:
\bea
\frac{W}{N_e} &=& \int dE\,dl\,\biggl\{
\dot\gamma P \left(kv_e l - \partial_l u \right)
\nnn
\\
&&{}
+\chi (\Gamma P - Y \rho)\ln\frac{e^{u/\chi} P(E)}{R(E)}\biggr\}
\eea
Since the sign of $\dot\gamma$ is arbitrary, and so is the shape of
$P(E,l)$, the first term must vanish pointwise, i.e.\ $\partial_l u =
k v_e l$. This of course makes sense: the stress contribution of each
element is just the strain derivative of its internal energy density. So
$u(E,l)-k v_e l^2/2$ is a function of $E$ only, and from the meaning of the
yield energy we expect this function to be $-E$ itself so that $u(E,l)=k v_e l^2/2 - E$.

With the flow term thereby eliminated,
\be
\frac{W}{N_e \chi Y} = \int dE\,dl\,\left(
\frac{\Gamma P}{Y} - \rho\right)\ln\frac{e^{u/\chi} P(E)}{R(E)}
\label{W_no_flow}
\ee
We now specialize to the conventional SGR model where
$\rho(E,l)=\rho(E)\delta(l)$, 
$\Gamma(E,l)=\Gamma_0 e^{u/\chi}$, and the obvious prior to use in the
definition of $S_\Lambda$ is $R(E)=\rho(E)$. Also,
the quantity 
\be
\pi(E,l)\equiv\Gamma(E,l) P(E,l)/Y
\ee
is a normalized distribution by
definition of $Y$. So we can express $W$ as
\bea
\frac{W}{N_e \chi Y} &=& \int dE\,dl\,
\pi(E,l) \ln\frac{\Gamma(E,l) P(E)}{\Gamma_0 \rho(E)}
\nnn
\\
&&{}- \int dE\,\rho(E)\ln\frac{e^{-E/\chi} P(E)}{\rho(E)}
\eea
To show that this is non-negative we now use a cross-entropy
inequality of the form 
\be 
\int dz\, P(z) \ln[P(z)/Q(z)] \geq -\ln\int dz\,
Q(z)
\label{crossent}
\ee 
This is valid if $P$ is a probability distribution and $Q$ is a
measure (i.e.\ a probability distribution but not necessarily
normalized to 1). The second term in $W$ is already of the form
(\ref{crossent}). In
the first term, we can get $\pi(E,l)$ inside the log by writing
$P(E)P(l|E)=P(E,l)$; here $P(l|E)$ is the conditional distribution of $l$ given
$E$. This yields
\bea
\frac{W}{N_e \chi Y} &=& \int dE\,dl\,
\pi(E,l) \ln\frac{Y\pi(E,l)}{\Gamma_0 P(l|E)\rho(E)}
\nnn
\\
&&{}
+ \int dE\,\rho(E)\ln\frac{\rho(E)}{e^{-E/\chi} P(E)}
\\
&\geq & -\ln(\Gamma_0/Y)-\ln\int dE\,e^{-E/\chi} P(E)
\\
&=& \ln \frac{\int dE\,dl\,e^{(k v_e l^2/2-E)/\chi} P(E,l)}
{\int dE\,dl\, e^{-E/\chi} P(E,l)} \geq 0
\eea
This shows that the SGR model is thermodynamically consistent in the
sense that its equation of motion ensures that the dissipation rate
$W$ is non-negative as it
should be. Note that this argument worked only because in writing down
the yield rate $\Gamma(E,l)$ above we had already identified the SGR
effective temperature $x$ with the thermodynamic temperature $\chi$ of
the slow degrees of freedom. This is an important conclusion, which we
will develop in more detail in the next Section.

We remark finally that the three inequalities used above to prove
$W\geq 0$ (two cross-entropy ones, and the last one that the average
of $\exp[k v_e l^2/(2\chi)]$ over any distribution is $\geq 1$) all
become equalities if (and only if) $P(E,l) \propto \rho(E) e^{E/\chi}
\delta(l)$. As is physically sensible, this is the (Boltzmann) steady
state of the 
SGR model without flow, where $W=0$. In all other situations, $W$ will
be positive.

\section{Constraints on more general SGR-like models}
\label{extensions}

The natural question arising from the results of the previous Section
is whether more general versions of the SGR equation of motion are thermodynamically consistent in the same sense.

\subsection{Deviations from local linear elasticity}

Consider first deviations from local linear elasticity, as
contemplated e.g.\ in the tensorial version of SGR~\cite{CatSol04}.
These are simple to incorporate: one could have $u(E,l)=u(l)-E$, with
the stress then the average of $u'(l)/v_e$ over all elements. As long
as $u(l)$ has its global minimum $u(0)=0$ at $l=0$, the argument above
goes through, and the model is thermodynamically consistent.

\subsection{More general yield rates $\Gamma(E,l)$}

One can also generalize the SGR model by allowing a more
general dependence of the yield rates $\Gamma(E,l)$ on the state of
the local element, i.e.\ $E$ and $l$. (This was done, for instance, in \cite{HeaAjdCat01} in the context of shear thickening models; we discuss the implications for that work in Section \ref{sec:conclusion}.) To understand the behaviour of
$W$ in this case, we start from the still general expression
(\ref{W_no_flow}), choosing again $R(E)=\rho(E)$ as the prior in
the entropy:
\bea
\frac{W}{N_e \chi Y} &=& \int dE\,dl\,
\pi(E,l) \ln\frac{e^{u/\chi} P(E)}{\rho(E)}
\nnn
\\
&&{}- \int dE\,\rho(E)\ln\frac{e^{-E/\chi} P(E)}{\rho(E)}
\eea
As before, one can now write
$P(E)=P(E,l)/P(l|E)=Y\pi(E,l)/[\Gamma(E,l)P(l|E)]$ so that the first
term becomes
\bea
&&\int dE\,dl\,
\pi(E,l) \ln\frac{e^{u/\chi} P(E)}{\rho(E)} =
\nnn
\\
&&=\ \int dE\,dl\,
\pi(E,l) \ln\frac{Y \pi(E,l)e^{u/\chi}}{\Gamma(E,l)P(l|E)\rho(E)}
\eea
Similarly, the second term is
\bea
&&\int dE\,\rho(E)\ln\frac{\rho(E)}{e^{-E/\chi} P(E)}
\nnn
\\
&&= \int dE\,dl\,P(l|E)\rho(E)\ln\frac{\Gamma(E,l)P(l|E)\rho(E)}{e^{-E/\chi}
Y\pi(E,l)}
\eea
Putting these two together and isolating the cross-entropy terms gives
\bea
\frac{W}{N_e \chi Y}
&=& \int dE\,dl\,
[\pi(E,l)-P(l|E)\rho(E)] \ln\frac{\pi(E,l)}{P(l|E)\rho(E)}
\nonumber\\
&&{}+ \int dE\,dl\,
\pi(E,l) \ln\frac{e^{u/\chi}}{\Gamma(E,l)}
\nonumber\\
&&{}
+ \int dE\,dl\,P(l|E)\rho(E)\ln\frac{\Gamma(E,l)}{e^{-E/\chi}}
\eea
To say something about necessary conditions on $\Gamma(E,0)$ first,
consider $P(l|E)=\delta(l)$, so that $\pi(E,l)\propto
\Gamma(E,0)P(E)\delta(l)$. Choose a $P(E)$ such that this is
$\rho(E)[1+\eta r(E)]\delta(l)$ with some small $\eta$ and a so far
unspecified function $r(E)$ satisfying $\int
dE\,\rho(E)r(E)=0$. Then the first contribution to $W$ is $O(\eta^2)$. The
second and third combine to
\be
-\eta \int dE\,\rho(E)r(E) \ln\frac{e^{-E/\chi}}{\Gamma(E,0)}
\ee
Unless the argument of the logarithm is a constant, one can arrange
the sign of $r(E)$ so as to make the integral positive, and thus 
$W$ negative to leading order in $\eta$. To avoid this contradiction
to $W\geq 0$, $\Gamma(E,0)$ must be proportional to $e^{-E/\chi}$, say
$\Gamma(E,0)=\Gamma_0 e^{-E/\chi}$. This argument shows that for
thermodynamic consistency, the effective temperature $x$ of SGR {\em
  must} be equal to the non-equilibrium thermodynamic temperature
$\chi$ of the configurational subsystem, as promised in the previous Section.

Turning next to $l$-dependence of $\Gamma(E,l)$, we can write this as
$\Gamma(E,l)=\Gamma_0 e^{-E/\chi}g(E,l)$ with $g(E,0)=1$, and hence
\bea
\frac{W}{N_e \chi Y}
&=& \int dE\,dl\,
[\pi(E,l)-P(l|E)\rho(E)] \ln\frac{\pi(E,l)}{P(l|E)\rho(E)}
\nonumber\\
&&{}+ \int dE\,dl\,
\pi(E,l) \ln\frac{e^{k v_e l^2/(2\chi)}}{g(E,l)}
\nonumber\\
&&{}+ \int dE\,dl\,P(l|E)\rho(E)\ln g(E,l)
\eea
The first term here is of cross-entropy form, while the second and third term are (pointwise) non-negative if $1\leq
g(E,l) \leq e^{k v_e l^2/(2\chi)}$. In summary, a sufficient condition
for $W\geq 0$ is
\be
\Gamma(E,l)=\Gamma_0 e^{-E/\chi}g(E,l)\ \mbox{with} \ 
1\leq g(E,l) \leq e^{k v_e l^2/(2\chi)}
\label{Gamma_conds}
\ee
The first part is also necessary as shown above. Whether the second
part is likewise necessary is not clear, i.e.\ there may be other
$g(E,l)$ that also keep $W\geq 0$. But physically the sufficient
condition given here is broad enough: the yield rate should increase
with $l^2$, i.e.\ it should never go below its $l=0$ value for given
$E$; on the other hand the yield rate should not go up more quickly
with $l^2$ than prescribed by the reduction in activation energy. If
anything, one might expect the rate to level off when the barrier
$-u=E-k v_e l^2$
becomes small or even negative, for example
\be
\Gamma(E,l)=\Gamma_0 \min(e^{u/\chi},1)
\ee
or
\be
\Gamma(E,l)=\Gamma_0 /(1+e^{-u/\chi})
\ee
It is clear from the above arguments that one could also allow a
shear-rate dependence of the attempt rate $\Gamma_0$, e.g.\ replacing 
$\Gamma_0$ by $(\Gamma_0^2+\dot\gamma^2)^{1/2}$ in the spirit of
recent proposals~\cite{WysMiyMatHuReiWei07} on how the relaxation
time varies with shear rate. These more general forms can of course
be combined with deviations from local linear elasticity as in the
previous subsection, with the upper limit on $g(E,l)$ then
$e^{u(l)/\chi}$ instead of $e^{k v_e l^2/(2\chi)}$.

\subsection{Frustration}

We had previously considered~\cite{Sollich98} the possibility that the
local strain $l$ might not relax fully to zero after a yield event
because disordered stresses from the environment prevent this, and
termed this effect `frustration'. Mathematically, it would correspond
to $\rho(E,l)=\rho(E)r(l|E)$ with $r(l|E)\neq \delta(l)$. For an SGR
model extended in this way it is doubtful that an argument can still
be made that $l$ should be left out of the entropy $S_\Lambda$ as $l$
no longer has deterministic dynamics slaved to the last yield time of
each element. If one nevertheless leaves $S_\Lambda$ as before, then
one can show that the extended model does not guarantee $W\geq 0$,
i.e.\ there are choices for $P(E,l)$ which lead to $W<0$. It is
possible that the model would still be thermodynamically consistent
because such distributions might not be accessible by the dynamics starting
from reasonable initial conditions, but we have not been able to
establish this.

\section{Summary and discussion}
\label{sec:conclusion}

With plausible expressions for the entropy and internal
energy of the $P(E,l)$-degrees of freedom, we have been able to show
that the SGR model admits an interpretation consistent with the
non-equilibrium thermodynamics framework of Bouchbinder and Langer. 
This interpretation requires that the
effective temperature $x$, postulated phenomenologically within the SGR model, is in fact the thermodynamic temperature $\chi$ of a
configurational subsystem describing mesoscopic degrees of freedom. (These degrees of freedom have slow energy transfer to the thermal bath of `fast modes' and hence their temperature is a nonequilibrium one.)

We believe this represents significant progress
in understanding the conceptual foundation of the SGR model. The original empirical motivation for introducing $x$ was (and remains) intuitively reasonable: an effective noise temperature arising ultimately from yield events throughout the sample. However a more solid formal grounding is now provided by the link to a thermodynamically defined
temperature that governs the exploration of the system's inherent
structures, through the dynamics of its mesoscopic degrees of freedom. These degrees of freedom are taken to be in thermal equilibrium with each other, but not with the rest of the system. If one accepts that, the thermodynamic interpretation gives new insights by showing, for example, that $x$ should be
viewed as intrinsic to the state of the system, and thereby to its history, as opposed to being set directly by the current rate of external driving. (The latter would make $x$-mediated effects second order, and hence negligible, within linear response theory~\cite{BouLan11b}.) 

Of course, the consistency of the thermodynamic interpretation of $x$ does not make this the only possible interpretation of the SGR model. It would be perfectly defensible to believe that, in many soft glasses, mesoscopic degrees of freedom do not equilibrate sufficiently {\em even among themselves} for their non-equilibrium temperature to ever be defined. In that case, however, it remains useful to know that the empirical choice made for the noise-driven dynamics in the standard formulation of SGR is exactly equivalent to having such a temperature. Anyone convinced that such a temperature does not exist would probably wish to add terms to break this equivalence -- which amounts to an implicit assumption of detailed balance among slow modes in the quiescent state. 

We have further shown that the thermodynamic consistency extends from
the original SGR model to more general versions that allow for
nonlinear local elasticity. Likewise, a fairly broad class of yield rates
$\Gamma(E,l)$ is allowed. This includes most physically plausible
choices, but always maintaining the thermally ($\chi$-)activated
dependence on yield energy.

Beyond checking consistency, the thermodynamic framework gives an
equation of motion (\ref{eqom}) for the entropy $S_C$ of the
configurational subsystem; see the Appendix for a generalization of this.
As discussed in~\cite{FalLan11}, one can
convert Eq.\ (\ref{eqom}) to an equation of motion for the effective
temperature (a more involved scenario is discussed
in~\cite{BouLan10}). To see this, recall the decomposition we
assumed for the energy of the slow subsystem, $U_C(S_C,\Lambda) =
U_\Lambda(\Lambda) + U_1(S_C-S_\Lambda(\Lambda))$. The effective temperature
$\chi=\left. \partial U_C/\partial S_C\right|_\Lambda =
\partial U_1/\partial S_1$ depends then on
$S_1=S_C-S_\Lambda$. Assuming as before that $S_\Lambda\ll S_C$
because $\Lambda$ only contains a vanishing fraction of the slow
degrees of freedom, $\chi$ becomes a function of $S_1\approx S_C$
only. Conversely, $\chi$ then fully determines $S_C$, so that $\dot
S_C = (\partial S_C/\partial \chi) \dot \chi$ and Eq.~(\ref{eqom}) becomes
\be
C_V^{\rm eff} \dot \chi = W + A(\theta-\chi)
\label{eqombis}
\ee
Here $C^V_{\rm eff}(\chi)$ is an effective heat capacity at constant volume,
and $A(\chi,\theta)$ is as before positive but
otherwise unspecified; in principle, $A$ could also depend
on other quantities like $\dot\gamma$ and possibly even $P(E,l)$. What
is interesting, however, is that the form of the driving term $W$ is
not negotiable; this is from (\ref{W_no_flow})
\bea
\frac{W}{N_e \chi} &=& \int dE\,dl\,
\Gamma(E,l) P(E,l) \ln\frac{e^{u/\chi} P(E)}{\rho(E)}
\nnn
\\
&&{}
- Y \int dE\,\rho(E)\ln\frac{e^{-E/\chi} P(E)}{\rho(E)}
\label{W_explicit}
\eea
While not obvious from this expression, it is clear from the definition
(\ref{W_def}) that in steady state, where $\Lambda\equiv P(E,l)$ does
not change, $W=V\sigma\dot\gamma$ is just the external work rate. This
is the form that was assumed recently in exploring the extension of
the SGR approach to understand shear banding: it was called `Model 1'
in~\cite{FieCatSol09}. Steady state flow curves of shear stress $\sigma$ versus
$\dot\gamma$ as predicted from (\ref{eqombis}) would therefore look
exactly like those found in~\cite{FieCatSol09}. On the other hand,
`Model 2' from that paper, where the driving term for $\chi$ is taken
as proportional to the yield rate $Y$, is not
thermodynamically consistent. We note that even with model 1, Eq.\ (\ref{W_explicit}) predicts different transient behaviour, because the driving term $W$  generally differs from
the external work rate $V\sigma\dot\gamma$. It would be interesting to
explore in future work what effects this thermodynamically motivated form for $W$ has on the predicted
shear banding dynamics.

We can similarly assess the thermodynamic consistency of two
extensions of the SGR model that were proposed some time ago to model shear
thickening effects~\cite{HeaAjdCat01}. That paper considered two models. In one, the effective
temperature $x\equiv \chi$ is taken as a function $x(\sigma)$ of the
overall stress. The second model, in a more radical departure from the standard SGR picture, takes $x(l)$ as a function of the local
strain in each element. The first case can feasibly be accommodated within the
thermodynamic framework, at least if we consider
situations where the effective temperature $x$, in its time evolution
according to (\ref{eqombis}), has reached a steady state. Assuming
that $\theta\ll x$ for simplicity, so that the effective temperature
is much larger than the bath temperature, the steady state condition
is $x=W/A=V\sigma\dot\gamma/A$. Given that $A$ is only restricted to
be positive but allowed to depend on e.g.\ $\dot\gamma$ and $\sigma$,
there are certainly choices for this dependence that would give a
steady state $x$ depending (only) on $\sigma$.

The second model in~\cite{HeaAjdCat01}, where $x=x(l)$, is different.
Having a temperature that differs from one element to
the next is a significant departure from the idea of a single
effective temperature for the slow configurational
subsystem. Consistent with this, one can check that a generically chosen $x(l)$ will violate the conditions~(\ref{Gamma_conds}). This does not of itself prove $W < 0$, but we have checked that there are indeed conditions under which negative $W$ does arise, e.g.\ for the specific (stepwise decreasing) form of $x(l)$ considered in~\cite{HeaAjdCat01}. Accordingly this second model of~\cite{HeaAjdCat01} is not consistent with a thermodynamic interpretation of the effective temperature as described here.

Finally, we should note again that our thermodynamic viewpoint for SGR was developed above by closely following the route first taken for STZ by Bouchbinder and Langer~\cite{BouLan09,BouLan09b,BouLan09c}. These authors used the second law to infer conditions on the dynamics of macroscopic variables. However these conditions, while both plausible and sufficient, are not in fact necessary: in the Appendix we describe a less restrictive, but still sufficient, set. The true value of thermodynamic consistency as a constraint on mesoscopic rheological models will not be known without a set of necessary, as well as sufficient, conditions for compliance of such models with the second law. It is possible that an explicitly model-independent formalism, such as that developed in~\cite{Oettinger05}, could prove useful in identifying such conditions. 

A further interesting direction for future work will be to explore
thermodynamic effects on the aging behaviour predicted by the SGR
model. An important feature of the model is that aging
effects are present~\cite{Sollich98,FieSolCat00} even when $\chi$ is
constant, as long as its value lies below the glass transition at
$\chi=E_0$. It is clear that allowing $\chi$ to evolve dynamically 
according to (\ref{eqombis}) in line with the thermodynamic picture
will contribute additional aging effects. However, what quantitative form these
take can only be fully assessed once physically motivated 
parameter dependences have been established for the effective heat
capacity $C_V^{\rm eff}$ and the coefficient $A$ determining the rate 
of heat transfer between the slow and fast subsystems.

{\bf Acknowledgements:} We thank Eran Bouchbinder, Mike Falk, Suzanne Fielding, Lisa Manning, and particularly Jim Langer, for illuminating discussions. We also thank KITP Santa Barbara, where this research was supported in part by the National Science Foundation under Grant No.\ NSF PHY05-51164. MEC is funded by the Royal Society.

\appendix
\section{More general conditions for second law}
\label{appendix}

We have in the main part of the paper adopted the argument of~\cite{BouLan09b} that the three
terms in (\ref{2nd_law}) must be separately non-negative. This is a
sufficient condition for the overall sum to be
non-negative and is also physically plausible. However it is not mathematically necessary. We discuss here a somewhat more general set of
conditions that are also sufficient. These admit the thermodynamic consistency of a wider class of models than allowed by the conditions of 
~\cite{BouLan09b} as used above.

If we write the first law as in (\ref{1st_law}) and insert $\dot U_R =
\theta_R \dot S_R$, we obtain
\be
\chi \dot S_C + \theta \dot S_K + \theta_R \dot S_R = W
\ee
This needs to be satisfied together with the second law
\be
\dot S_C + \dot S_K + \dot S_R \geq 0
\ee
With the reversible degrees of freedom ($\Lambda$) of the slow system
already treated separately in $W$, these two equations are completely
symmetric in the roles played by the slow, fast and reservoir
subsystems. A more general set of conditions for the second law to
hold then suggests itself as $W\geq 0$ and
\bea
\chi \dot S_C &=& \alpha_C W + A(\theta-\chi) +
C(\theta_R-\chi)
\\
\theta \dot S_K &=& \alpha_K W + A(\chi-\theta) +
B(\theta_R-\theta)
\\
\theta_R \dot S_R &=& \alpha_R W + 
B(\theta-\theta_R) + C(\chi-\theta_R)
\eea
where $A$, $B$, $C$, $\alpha_C$, $\alpha_K$, $\alpha_R$ are all
non-negative and $\alpha_C+\alpha_K+\alpha_R=1$. The last condition
is enough to ensure that the first law is satisfied, as the heat flow
terms all cancel in pairs. For the total rate of entropy change one gets
\bea
\dot S_C + \dot S_K + \dot S_R &=&
W\left(\frac{\alpha_C}{\chi}+\frac{\alpha_K}{\theta}
+\frac{\alpha_R}{\theta_R}\right)
+A\,\frac{(\chi-\theta)^2}{\chi\theta}
\nnn
\\
&&{}
+B\,\frac{(\theta-\theta_R)^2}{\theta\theta_R}
+C\,\frac{(\chi-\theta_R)^2}{\chi\theta_R}
\eea
which is evidently non-negative. The more restrictive condition of~\cite{BouLan09b} corresponds essentially to $\alpha_K=\alpha_R=C=0$. One has then
$\dot U_R = \theta_R \dot S_R = -B(\theta_R-\theta)$ as in
(\ref{inequalities}), and
\be
\theta\dot S_K + \frac{\theta}{\theta_R} \dot U_R
= A(\chi-\theta) + \frac{B}{\theta_R}(\theta_R-\theta)^2
\ee
This generalizes the second equality of (\ref{inequalities}) to the
case where $\theta\neq \theta_R$. In the limit where $B$ becomes large
at constant $B(\theta_R-\theta)$, the term on the r.h.s.\ proportional
to $B$ goes to zero so (\ref{inequalities}) is retrieved. That a term
proportional to $B(\theta_R-\theta)^2$ should be present in general
here is clear because $\theta \dot S_K + (\theta/\theta_R)\dot U_R =
\theta(\dot S_K + \dot S_R)$ and this rate of entropy change should
have a non-zero contribution when there is heat transfer at a finite
rate between the reservoir and the kinetic-vibrational subsystem.

Returning to our more general conditions above, the effect of the
coefficient $C$ is probably unimportant so long as we are indeed interested only in
the limit $\theta=\theta_R$ because of strong thermal coupling to
the reservoir: one can then just combine $A$ and $C$.

The effect of the $\alpha$ coefficients is more subtle. In soft
materials, where the reservoir should include the solvent degrees of
freedom, it seems plausible that a substantial amount of the work
performed on the system (minus the amount stored reversibly in
$\Lambda$, which gives $W$) could be dissipated directly in the
solvent. For instance, in a dense emulsion, work could be dissipated directly in the solvent films between droplets without ever passing through the $\Lambda$ degrees of freedom. In soft glasses, therefore, $\alpha_R$ could be substantial, and correspondingly
$\alpha_C<1$. An alternative strategy might be to exclude this contribution from the work input, by separating off a global Newtonian solvent contribution. However, as the example of thin films between emulsion droplets shows, the mesoscopic and solvent contributions cannot be treated as simply additive: in principle the droplet organization controls both the thicknesses of films and the shear rates within them. It therefore seems advisable not to assume that kind of separation within a thermodynamic description, at least in the case of soft glasses.

Despite the above generalizations the condition that $W\geq 0$ remains in place. So far we have not been able to show
mathematically that $W\ge 0$ is necessary (or indeed that there are not
broader conditions on the heat flows that would satisfy the second
law), although physically it is certainly plausible that the dissipation
rate should be non-negative. The main possible generalization
arising from the arguments above is then in the equation of motion for
$\chi$, which would become (replacing (\ref{eqombis}))
\be
C_V^{\rm eff} \dot \chi = \alpha_C W + A(\theta-\chi)
\ee
Here one now has an additional undetermined factor $0\leq \alpha_C\leq
1$ in front of the driving term $W$. One could speculate whether there
is a difference between e.g.\ metallic and soft glasses in the size of
$\alpha_C$, which could be of order unity for the former and small for
the latter.

\bibliography{/home/psollich/references/references,notes}

\end{document}